\documentclass[12pt,tightenlines,nofootinbib]{revtex4}
\usepackage{url}
\usepackage{bbm}
\usepackage{mathrsfs}
\usepackage{epsfig,psfrag}
\usepackage{amsmath,amsfonts,amssymb}
\usepackage{epigraph}
\usepackage{bm}
\usepackage{physics}

\newcommand{\bfx}{{\bf x}}

\newcommand{\bfn}{{\bf n}}

\newcommand{\bfE}{{\bf E}}
\newcommand{\bfH}{{\bf H}}
\newcommand{\bfA}{{\bf A}}
\newcommand{\bfB}{{\bf B}}

\newcommand{\bfK}{{\bf K}}

\newcommand{\bfP}{{\bf P}}

\newcommand{\cF}{{\mathcal F}}
 
\newcommand{\cS}{{\mathcal S}}
\newcommand{\cD}{{\mathcal D}}

\newcommand{\be}{\begin{equation}}
\newcommand{\ee}{\end{equation}}

\newcommand{\cev}[1]{\reflectbox{\ensuremath{\vec{\reflectbox{\ensuremath{#1}}}}}}

\newcommand{\cZ}{\mathcal Z}

\def\Tr{\,{\rm Tr}\,}
\def\tr{\,{\rm tr}\,}

\newcommand{\bea}{\begin{eqnarray}}
\newcommand{\eea}{\end{eqnarray}}
\newcommand{\ben}{\begin{enumerate}}
\newcommand{\een}{\end{enumerate}}
\newcommand{\bit}{\begin{itemize}}
\newcommand{\eit}{\end{itemize}}

\def\k{\kappa}

\newcommand{\cbA}{\bm{\mathcal{A}}}

\begin{document}

\markboth{
Giuseppe Bimonte, Thorsten Emig, Noah Graham, and Mehran Kardar
}{Surface Approaches to Quantum Fluctuations and the Casimir Force}

\title{
Something Can Come of Nothing:  
Surface Approaches to
Quantum Fluctuations and the Casimir Force}

\author{Giuseppe Bimonte}
\affiliation{Dipartimento di Fisica E. Pancini, Università di Napoli Federico II, Complesso Universitario di Monte S. Angelo, Via Cintia, I-80126 Napoli, Italy}
\affiliation{INFN Sezione di Napoli, I-80126 Napoli, Italy}

\author{Thorsten Emig}
\affiliation{Laboratoire de Physique Théorique et Modèles Statistiques, CNRS UMR 8626, Université Paris-Saclay, 91405 Orsay CEDEX, France}

\author{Noah Graham}
\affiliation{Department of Physics, Middlebury College,
Middlebury, VT  05753, USA} 

\author{Mehran Kardar}
\affiliation{Department of Physics, Massachusetts Institute of
Technology, Cambridge, MA 02139, USA}

%Abstract
\begin{abstract}
The Casimir force provides a striking example of the effects of quantum fluctuations in a mesoscopic system.  Because it arises from the objects' electromagnetic response, the necessary calculations in quantum field theory are most naturally expressed in terms of electromagnetic scattering from each object.  In this review we illustrate a variety of such techniques, with a focus on those that can be expressed in terms of surface effects, including both idealized boundary conditions and their physical realization in terms of material properties.
\end{abstract}

%Keywords, etc.
%\begin{keywords}
%casimir forces, fluctuation induced interactions, stress tensor, path integral, scattering matrix
%\end{keywords}

\maketitle

\tableofcontents

\epigraph{Nothing can come of nothing.}
{\textit{King Lear} (I, i, 92)}

\epigraph{It's cool, huh? Zero-point energy. I save the best
inventions for myself.}{Syndrome, \textit{The Incredibles}}

\section{Introduction:  The Quantum Vacuum}
Since it was first calculated in 1948 \cite{Casimir48-2}, the Casimir force between uncharged conducting plates has provided a compelling application of quantum field theory.  Proportional to both $\hbar$ and $c$, it provides a concrete illustration of the effects of relativistic quantum fluctuations in a mesoscopic system.  Since then, it has evolved from thought experiment to precision measurement \cite{Lamoreaux97,Mohideen98,Roy99,Ederth00,Chan01,Chen02,DeKiewiet03,Harber05,Chen06,Krause07,Decca07,Chen07,Munday07,Chan08,Kim08,Palasantzas08,Munday09}, and may in the future guide and influence the design and operation of microelectromechanical devices \cite{Chan01,Capasso07}.

A standard calculation in undergraduate electromagnetism shows that one can compute the total energy of an electrostatic charge distribution as an integral over space of either the electrostatic interaction energy of all the charges, or the energy density in the electric field,
\begin{equation}
U = \frac{1}{2}  \int \phi \rho \, d{\bf x} = 
-\frac{\epsilon_0}{2} \int \phi {\bf\nabla}^2 \phi \, d{\bf x}  =
\frac{\epsilon_0}{2}  \int {\bf\nabla} \phi \cdot {\bf\nabla} \phi \, d{\bf x} =\frac{\epsilon_0}{2} \int \bm{E}^2\, d{\bf x} \,,
\end{equation}
where $\bm{E}$ is the electric field, $\phi$ is the electrostatic potential,
and $\rho$ is the charge density; the contributions from the surface at infinity are assumed to vanish.  Because this derivation requires an integration by parts, it does not imply that the integrands are equal, and it is only the latter approach that yields an energy density consistent with the locality constraints of relativity.  From a theoretical point of view, this argument yields the remarkable result that the energy of a parallel-plate capacitor is stored not on the plates themselves, but in the electric field between the plates.  However, since any experiments not involving gravitation are sensitive only to the total energy of a system rather than its spatial distribution within that system, in practice both approaches yield equivalent results.  Furthermore, by writing the electrostatic potential in terms of the Green's function $G({\bf x},{\bf x'})$, one can express the energy as an interaction between charges,
\begin{equation}
U = \frac{1}{2}  \int \phi \rho \, d{\bf x} = 
    \frac{1}{8 \pi \epsilon_0}  \int \int \rho({\bf x}) G({\bf x},{\bf x'})
    \rho({\bf x'}) \, d{\bf x} d{\bf x'} \,.
\end{equation}

An  exactly analogous situation applies to quantum fluctuations in quantum electrodynamics (QED).  As a QED phenomenon, the Casimir energy can be represented either in terms of fluctuating charges and currents or fluctuating fields (and in scattering approaches it is often convenient computationally to pass between the two representations \cite{PhysRevD.80.085021}), but it is no more or less a demonstration of the existence of ``vacuum fluctuations'' than any other QED process \cite{PhysRevD.72.021301}.  But once one considers gravitational phenomena, these choices become inequivalent, and one must consider the field interpretation to maintain consistency with the equivalence principle \cite{PhysRevD.89.064027}.

These gravitational effects play a particularly striking role in cosmological ``dark energy.''  In any QED calculation, whether it might be mesoscopic Casimir system, an atomic Lamb shift, or a high-energy physics scattering amplitude, one encounters a formally infinite energy corresponding to the zero-point oscillation $\frac{1}{2} \hbar \omega$ of all the modes of each quantum field.  This infinity can be regularized, so that it is rendered finite by assuming that the quantum field theory is modified at very short distances by new physics that acts as a cutoff on these integrals.  However, since that cutoff must be at much shorter distances that those probed by current experiments, the resulting integral would yield an extremely large value, corresponding to an energy density per unit volume of the vacuum.  In any measurement of Casimir forces, this large energy cancels and has no observable consequences.

Since the dynamics of the expanding universe are sensitive to absolute sources of energy rather than just energy differences, this vacuum energy has major effects in cosmology.  These effects are distinctive because a fixed vacuum energy per unit volume has negative pressure: as the universe expands, the total energy goes up, in contrast to the usual situation in thermodynamics where an increase in volume corresponds to a decrease in energy.  So vacuum energy leads to a relation $dU = -p dV$ with negative pressure $p$, corresponding  to a stress-energy tensor that is proportional to the spacetime metric.  Because this negative pressure appears in three entries of the stress-energy tensor while the positive energy appears only once, the resulting effect on the expansion rate is the opposite of ordinary matter --- dark energy accelerates the expansion of the universe, in contrast to the ordinary forces of gravitational attraction that slow it down.  The existence of our observed universe, however, shows that the actual vacuum energy must be much smaller than one would estimate from the sum over zero-point energies --- otherwise this accelerated expansion would have prevented the formation of stars and galaxies.  At first it was assumed that some unknown mechanism instead canceled their contribution to the gravitational stress-energy tensor.  However, ``just because it's infinity doesn't mean it's zero'':  Precision cosmology experiments \cite{SupernovaSearchTeam:1998fmf,SupernovaCosmologyProject:1998vns} have offered convincing evidence that there is a nonzero vacuum energy density, just one that is 120 orders of magnitude smaller than what would expect from an estimate based on the integral over zero-point modes with a short-distance cutoff.

Since Fermi fields obey canonical anticommutation rather than commutation relations, each mode of a Fermi field has zero point energy $-\frac{1}{2} \hbar \omega$ rather than $+\frac{1}{2} \hbar \omega$.  As a result, one might hope that this vacuum energy cancels between Bose and Fermi fields.  This proposal is made concrete in theories with supersymmetry, in which every fermionic field has a bosonic partner and vice versa, guaranteeing that the vacuum energy vanishes.  However, since supersymmetry is not observed in nature --- there is no massless fermionic partner to the photon, for example --- supersymmetry must be spontaneously broken, which re-introduces a nonzero vacuum energy density at the associated energy scale.  While the resulting vacuum energy density is much smaller than that of the short-distance cutoff, it is still far, far larger than the observed value; since no experimental evidence for supersymmetry has been observed, the supersymmetry breaking scale must be well above the scales that have been probed in high energy physics.

It is important to emphasize that this mystery does not represent a physical or logical inconsistency in the observed energy of the vacuum, however.  Like a mass or coupling constant, the vacuum energy is a renormalized parameter in quantum field theory, meaning that it can be fixed to whatever value is dictated by experimental input via the corresponding counterterm in the theory.  The ``bare'' parameter alone is not physically measurable.  In the case of a mass or coupling, however, one can fix the renormalized parameter on the basis of one experiment and then use that value in many other experiments, giving the choice predictive power.  For the energy of the vacuum, in effect we have one experiment for one free parameter.  As a result, one is motivated to find more sophisticated models in which this parameter can be connected to other properties of the theory that can be fixed independently, using as guidance approaches such as the anthropic principle \cite{PhysRevLett.59.2607}, or nonlinear feedback mechanisms such as quintessence \cite{Tsujikawa_2013} or effective media \cite{Leonhardt_2019}, but how to do so precisely remains an open question.

\section{Proximity force approximation and derivative expansion}

As is typical in classical and quantum field theory, exact analytic calculations are only tractable in problems with a high degree of symmetry.  It is therefore invaluable, especially in comparison with experimental work, to have convenient approximations available.  

Originally developed by Derjaguin~\cite{Derjaguin1934}  in the context of surface adhesion and colloids, the proximity force approximation (PFA) relates forces between gently curved objects at close separations, $\sim d$, to the corresponding interactions between flat surfaces over an area set by the local radii of curvature, $\sim R$.
In this approach, one approximates the Casimir energy between two surfaces as the local interaction between flat parallel plates
\begin{equation}
E_{\rm PFA}=\int_{\Sigma} d{\bf x}\,U(z) \,,
\end{equation}
where $z$ is the local distance between the surfaces and $U(z)$ is the energy per unit area for parallel plates made of the same materials, which is given below in Eq.~(\ref{eq:CasimirDensity}) for the case of ideal boundaries.  Here the integral is over the surface $\Sigma$, which can be taken as one of the two interacting surfaces or as a reference surface placed in between them.  Since this approximation ignores the effects of non-parallelism of the interacting surfaces, it can depend on the choice of $\Sigma$, meaning that a different choice may lead to a better or worse approximation.

PFA is widely used to estimate Casimir~\cite{Casimir48-2} and van der Waals~\cite{Parsegian} forces, starting with the Lifshitz formula~\cite{Lifshitz} for the interaction between parallel plates, and is asymptotically exact as $d/R\to0$.  However, the range of validity of this approximation, and the nature of subsequent correction terms to PFA, were previously not known.  A conceptual breakthrough in 2011~\cite{PhysRevD.84.105031} showed that earlier perturbative corrections to parallel plate forces \cite{PhysRevLett.87.260402,Neto} can be organized into a gradient expansion in the local separation between surfaces for the force between gently curved bodies, implementing this program for scalar fields. Recognizing the usefulness of this approach, Ref.~\cite{Bimonte_2012} generalized the gradient expansion of Ref.~\cite{PhysRevD.84.105031} to practical computations of the electromagnetic Casimir force and tested the validity of the PFA corrections against exact results for the force between a perfectly conducting sphere and plate.  The extension to real materials described by a (frequency-dependent) dielectric response $\epsilon(\omega)$ is not trivial, and was developed in Ref.~\cite{MaterialGradient}.  In particular, this work obtained an explicit expression for the PFA correction to the force gradient using the dielectric function of gold, at room temperature.  The corrections to the Casimir free energy were found to scale logarithmically with distance, with an unexpectedly large temperature dependence.  Further work~\cite{RoughHeat} considered the effects of roughness or surface modulations.  By now the gradient expansion has been confirmed against a number of exact results for spheres and cylinders~\cite{Teo_2013}, and in experiments with corrugated surfaces~\cite{GradientX}.

\subsection{The gradient expansion from resummation of perturbation theory}

Consider two bodies with gently curved surfaces described by (single-valued) height profiles $z=H_1({\bf x})$ and $z=H_2({\bf x})$, with respect to a reference plane $\Sigma$, where ${\bf x} \equiv (x,y)$ are Cartesian coordinates on $\Sigma$ and the $z$ axis is normal to $\Sigma$.  The intervening quantum field can be a scalar or electromagnetic (EM) field.  For the scalar case we can impose either Dirichlet (D) or Neumann (N) boundary conditions.  The EM case, for ideal (mirror) boundaries, is obtained simply as the sum of D and N cases, corresponding to the two transverse polarizations.  Extension to the case of real materials, described by a complex dielectric permittivity, is possible.  The only restriction on the boundary conditions is that they should describe homogeneous and isotropic materials, so that the {Casimir} energy is invariant under simultaneous translations and rotations of the two profiles in the plane $\Sigma$.

The gradient expansion postulates that the Casimir energy, when generalized to two surfaces and to arbitrary fields subject to arbitrary boundary conditions, is a functional $E[H_1,H_2]$ of the heights $H_1$ and $H_2$, which has a derivative expansion
\begin{eqnarray}\label{2surfaces}
E[H_1,H_2]=\int_{\Sigma} d{\bf x}\,U(H)\, &&\left[1+\beta_1 (H)
{\bf \nabla} H_1\cdot {\bf\nabla} H_1 \right.  \nonumber\\
&&+\left.\beta_2 (H) {\bf\nabla} H_2\cdot{\bf\nabla} H_2 +
\beta_\times (H) {\bf\nabla} H_1\cdot{\bf\nabla} H_2
\right.
\nonumber\\&&
+ \left.\beta_{\rm -} (H)\, {\hat {\bf z}} \cdot
({\bf\nabla} H_1 {\bf \times} {\bf\nabla} H_2) + \cdots\right],
\end{eqnarray}
where $H({\bf x}) \equiv H_2({\bf x})-H_1({\bf x})$ is the height difference, and dots denote higher derivative terms.  Here, $U(H)$ is the energy per unit area between parallel plates at separation $H$; translation and rotation symmetries in ${\bf x}$ only permit four distinct gradient coefficients at lowest order, $\beta_1 (H)$, $\beta_2 (H)$, $\beta_\times (H)$ and $\beta_- (H)$ .  The form of such a local expansion is motivated by the existence of well-established derivative expansions of scattering amplitudes~\cite{Voronovich_1994} from which the Casimir energy can be derived~\cite{PhysRevD.80.085021}.  Arbitrariness in the choice of $\Sigma$ constrains the coefficients $\beta$ in the above expansion.  Invariance of $E$ under a parallel displacement of $\Sigma$ requires that all the $\beta$'s depend only on the height difference $H$, and not on the individual heights $H_1$ and $H_2$.  These coefficients are further constrained by the invariance of $E$ with respect to tilting the reference plate $\Sigma$.  Under a tilt of $\Sigma$ by an infinitesimal angle $\epsilon$ in the $(x,z)$ plane, the height profiles $H_i$ undergo a change  by $\Delta H_i = -\epsilon\left(x+H_i \frac{\partial H_i}{\partial x}\right)$, and the requirement that  $E$ not change implies 
\begin{eqnarray}\label{tilt}
&&2\left(\beta_1(H)+\beta_2(H)\right)+2\,\beta_{\times}(H)+
\,H\,\frac{d \log U}{d H}-1=0\;,\nonumber\\
&&\beta_{\rm -}(H)=0\;,
\label{betarelations}
\end{eqnarray}
so that the non-vanishing cross term $\beta_{\times}$ is determined by $\beta_1$, $\beta_2$ and $U$.

Equations~(\ref{betarelations}) indicate that, to second order in the gradient expansion, the two-surface problem reduces to that of a single curved surface facing a plane. Therefore, $U$, $\beta_{1}$ and $\beta_{2}$ can be determined by setting  $H_1$ or $H_2$ to be zero.  Let us set $H_1=0$ and define $\beta_{2}(H)\equiv\beta(H)$.  We can then determine the exact functional dependence of $\beta(H)$ on $H$ by comparing the gradient expansion Eq.\ (\ref{2surfaces}) to a perturbative expansion of the Casimir energy around flat plates, to second order in the deformation.  For this purpose, we take $\Sigma$ to be a planar surface and decompose the height of the curved surface as $H({\bf x})=d+h({\bf x})$, where $d$ is chosen to be the distance of closest separation.  For small deformations $|h({\bf x})|/d \ll 1$ we can expand $E[0,d+h]$ as
\begin{equation}
E[0,d+h]=A\,U(d)+\,\mu(d) \tilde h({\bf 0}) +  
\,\int \frac{d{\bf k}}{(2 \pi)^2} G({ k};d)|{\tilde h}({\bf k})|^2\;,
\end{equation}
where $A$ is the area, ${\bf k}$ is the in-plane wave-vector and ${\tilde h}({\bf k})$ is the Fourier transform of $h({\bf x})$.  The kernel $ G({k};d)$ has been evaluated by several authors: in Ref.~\cite{PhysRevLett.87.260402} for a scalar field fulfilling D or N boundary conditions on both plates, as well for the EM field satisfying ideal metal boundary conditions on both plates; and in Ref.~\cite{Neto} for the EM field with dielectric boundary conditions.  For a deformation with small slope, the Fourier transform ${\tilde h}({\bf k})$ is peaked around zero.  Since the kernel can be expanded at least through order $k^{2}$ about $k=0$~\cite{Voronovich_1994}, we define
\begin{equation}
G({k};d)=\gamma(d)+\delta(d)\,k^2 \,+\cdots\;.
\end{equation}
For small $h$, the coefficients in the derivative expansion can be matched with the perturbative result.  By expanding Eq.~(\ref{2surfaces}) in powers of $h({\bf x})$ and comparing the result with the perturbative expansion to second order in both ${\tilde h}({\bf k})$ and $k^2$, we obtain 
 \begin{equation}\label{eq:PFAsum}
  U'(d)= \mu(d) \, ,\quad U''(d)=2 \gamma(d) \, ,
 \quad \beta(d)=\frac{\delta(d)}{U(d)}\;,
\end{equation}
with prime denoting a derivative.  

\subsection{Ideal boundaries}
For ideal boundary conditions (D or N for a scalar field, or perfect mirrors for the EM field) the energy per unit area between parallel plates with separation $d$ is given by
\begin{equation}\label{eq:CasimirDensity}
U(d) = -  \alpha \frac{\pi^2 \hbar c}{1440 d^3}\,.
\end{equation}
The overall coefficient is $\alpha = 1$ for scalar field fluctuations where both surfaces have either Dirichlet or Neumann boundary conditions, $\alpha = 2$ (corresponding to the two polarizations) for electromagnetic fluctuations with perfect conductor boundary conditions, and $\alpha = -7/8$ for a scalar field with mixed Neumann/Dirichlet boundaries.  

Using the Eqs.~\ref{eq:PFAsum}, and appropriate perturbation series, it is then possible to compute the coefficient $\beta$ for the following five cases: a scalar field obeying D or N boundary conditions on both surfaces; D boundary conditions on the curved surface and N boundary conditions on the flat surface, or {\it vice versa}; and for the EM field with ideal metal boundary conditions.  Since in all these cases the problem involves no other length apart from the separation $d$, $\beta$ is a pure number.  $\beta_{\rm D}$ was first computed in~\cite{PhysRevD.84.105031}, and found to be $\beta_{\rm D}=2/3$.  
Ref.~\cite{Bimonte_2012} then found find $\beta_{\rm N}=2/3 \,(1-30/\pi^2)$,   $\beta_{\rm DN}=2/3$, $\beta_{\rm ND}=2/3-80/7\pi^{2}$ (where the double subscripts denote the curved surface and flat surface boundary conditions respectively), and $\beta_{\rm EM}=2/3 \,(1-15/\pi^2)$.  Upon solving Eqs.~(\ref{tilt})  one then finds $\beta_{\times}=2 - \beta_1 -\beta_2$, where $\beta_1$ and $\beta_2$ are chosen to be both equal to either $\beta_{\rm D}$, $\beta_{\rm N}$ or $\beta_{\rm EM}$, for the case of identical boundary conditions on the two surfaces, or rather $\beta_1=\beta_{\rm DN}$ and $\beta_2=\beta_{\rm ND}$ for the case of a scalar field obeying mixed ND boundary conditions. 

Using the above values for $\beta$ and $\beta_{\times}$, it is possible to find the leading correction to PFA by explicit evaluation of Eq.~(\ref{2surfaces}) for the desired profiles.  For example, for two spheres of radii $R_1$ and $R_2$,  both with the same boundary conditions for simplicity, we obtain:
\begin{equation}
\label{eq:2_spheres}
E=E_{\rm PFA} \left[1- \frac{d}{R_1+R_2}+(2 \beta-1)\left(\frac{d}{R_1}+\frac{d}{R_2} \right)
\right]\;,\end{equation} 
where $E_{\rm PFA}=-(\alpha \pi^3 \hbar c R_1 R_2))/[{1440 d^2 (R_1+R_2)}]$.  The corresponding formula for the sphere-plane case can be obtained by taking one of the two radii to infinity. These results are in good agreement with analytic calculations in the sphere-plane system~\cite{Teo_2011,Teo_2013}.

\subsection{Material boundaries}

The approach described above, converting a perturbative expansion in small deformations to a gradient expansion, can be performed for any cases where a perturbative expansion is possible.  These include the case of two infinitely thick plates separated by $d$, each modeled as a homogeneous and isotropic dielectric material, with frequency dependent permittivities $\epsilon_1(\omega)$ and $\epsilon_2(\omega)$.  Much like the crossover between retarded and non-retarded van der Waals interactions, the Casimir force also becomes dependent on material properties through $\epsilon(\omega)$ at short distances (typically in the range of 10-100~nm).  If the dielectric response is dominated by a resonance at a single characteristic frequency $\overline{\omega}$, in the so-called near-field regime of separations $d\leq c/\overline{\omega}$, the Casimir energy density in Eq.~(\ref{eq:CasimirDensity}) changes from $U(d)\propto \hbar c/d^3$ to $U(d)\propto \hbar\overline{\omega}/d^2$.

Since $U(d)$ still diverges as $d\to0$ even in the near field regime, the PFA and gradient expansion should still be applicable.  This calculation was implemented in Ref.~\cite{MaterialGradient}, which applied this method to two infinitely thick plates composed of homogeneous and isotropic dielectric materials, with general permittivities $\epsilon_1(\omega)$ and $\epsilon_2(\omega)$.  The resulting corrections to PFA are no longer pure numbers, but depend on temperature and separation (through material dependent parameters).  While the details will not be reproduced here, we note that they have in fact been used for comparison to sphere/plate experiments~\cite{GradientX,WOS:000643632900001,WOS:000449784800006}.

In principle, we may anticipate a gradient expansion for any situation when a shape dependent quantity diverges sufficiently rapidly as a function of separation.  One example, described in Ref.~\cite{RoughHeat}, involves radiative heat transfer (RHT) between objects at different temperatures.  The heat transfer between parallel plates at large distances is dominated by propagating photons, and is independent of separation $d$.  However, as noted by Polder and Van Hove~\cite{VanHove}, RHT at short separation increases strongly upon decreasing separation due to tunneling of evanescent EM waves across the vacuum gap.  The precise form of this enhancement depends on the material properties: If the response of the material can be characterized by single dominant frequency $\overline{\omega}$, in the  near-field regime of  $d\leq c/\overline{\omega}$, RHT diverges as $S(d)\propto \hbar\overline{\omega}^2/d^2$~\cite{Golyk_2013}.  RHT due to evanescent waves has also attracted a lot of interest due to its connection with scanning tunneling microscopy and scanning thermal microscopy under ultra-high vacuum conditions~\cite{Pendry,doi:10.1146/annurev.matsci.29.1.505}.  The enhancement of HT in the near-field regime (generally denoting separations small compared to the thermal wavelength, which is roughly 8 microns at room temperature) has only recently been verified experimentally~\cite{exp1,exp2}. 

Much like PFA, the leading term for RHT between closely spaced curved objects has been computed by use of a corresponding proximity transfer approximation (PTA)~\cite{scattering2,PTA1,Fan}.  Ref.~\cite{Golyk_2013} extended a gradient expansion approach to compute the first correction to PTA, which can be expressed as a spectral decomposition, with a correction factor $\beta(\omega)$ at each frequency.

\subsection{Further applications}

While the above examples focused on two surfaces, the gradient expansion can also be applied to estimate the interaction between a polarizable particle and a gently curved surface.  For a particle of polarizability $\alpha$, the Casimir-Polder force~\cite{Casimir48-1} due to a flat surface scales with the separation $d$ as $\alpha\times \hbar c/d^4$ in the far-field, and as $\alpha\times \hbar \overline{\omega}/d^3$ in the near-field.  Perturbative results for the interaction of the particle with a slightly modulated surface can again be converted to a gradient expansion for the force close to a curved surface.  In Ref.~\cite{PhysRevD.90.081702} the leading correction on approaching a surface with profile $H({\bf x})$ is shown to depend on its curvature (proportional to $d\nabla^2 H$), while the next order scales as $[d\nabla^2 H]^2$.  If the polarizable particle is a molecule, Ref.~\cite{PhysRevA.94.022509} demonstrates that in principle the shift of its rotational levels due to Casimir-Polder interactions can be used as a probe of the surface profile.

\section{The scattering approach and its applications}

The calculations we have studied in the previous section are based on local approximations.  Scattering methods provide a complementary approach, based on the global electromagnetic response of each object in a Fourier scattering basis.  As a result, this approach provides exact expressions that are applicable both to concrete calculations and broader theoretical analysis.

The fundamental idea of the scattering approach is to decompose the combined quantum fluctuations of a pair of objects in terms of the fluctuations of each object individually, combined with a propagator that carries these fluctuations from one object to another.  The result is the so-called ``TGTG'' form combining the scattering $T$-matrix \cite{Waterman65,Waterman71} for each object with the free Green's function propagator \cite{PhysRevLett.97.160401}.  The material and geometric properties of each object are represented through its individual $T$-matrix, independent of the other objects, while the propagator captures information about the objects' relative position and orientation but is independent of the properties of the objects themselves.  Since the scattering matrices for different objects are often most naturally written in different bases --- for example, spherical and Cartesian coordinates for a sphere-plane system --- the relevant change of basis may also be needed in combining these expressions.  Finally, the sum over all possible fluctuations is introduced as a log-determinant or equivalently a trace-log, which in practice is expressed as an integral over frequencies and a sum over scattering channels, with the former typically Wick rotated to the imaginary frequency axis.

Scattering theory methods were first applied to the parallel plate geometry, by reformulating Lifshitz theory in terms of reflection coefficients \cite{Kats77}, leading to a derivation of  the Lifshitz formula using reflection coefficients for lossless infinite plates \cite{Jaekel91} and its extensions to lossy \cite{Genet03} and non-specular reflection  \cite{Lambrecht06}.  Around the same time, the multiple scattering approach to the Casimir energy for perfect metal objects was developed, making it possible to compute the Casimir energy at asymptotically large separations \cite{Balian77,Balian78} at both zero and nonzero temperature. In this approach, information about the conductors is encoded in a local surface scattering kernel.

Another scattering-based approach has been to express the Casimir energy as an integral over the density of states of the fluctuating field, using the Krein formula \cite{Krein53,Krein62,Birman62} to relate the density of states $\rho(k)$ to the $\cS$-matrix for scattering from the ensemble of objects,
\begin{equation}
\rho(k)  =  \frac{1}{2 \pi i} \Tr \frac{d}{dk} \log \cS(k)\,.
\end{equation}
This $\cS$-matrix is difficult to compute in general, but the use of many-body scattering of scalar fields made it possible to connect the $\cS$-matrix of a collection of spheres \cite{Henseler97} or disks \cite{Wirzba99} to the objects' individual $\cS$-matrices, which are easy to find. Subsequent work combined this result with the Krein formula to investigate the scalar and fermionic Casimir effect for disks and spheres \cite{Bulgac01,Bulgac06,Wirzba08}. Casimir energies of solitons in renormalizable quantum field theories have been computed using scattering theory techniques that combine analytic and numerical methods \cite{Graham09}.

Refs. \cite{Bordag85,Robaschik87} introduced path integral methods to the study of Casimir effects and used them to investigate the electromagnetic Casimir effect for two parallel perfect metal plates.  Similar methods were used to study the scalar thermal Casimir effect for Dirichlet, Neumann, and mixed boundary conditions in Refs.\ \cite{Li91,Li92}. This approach was adapted to the quantum case and developed further in Refs.\ \cite{Golestanian97,Golestanian98} and was subsequently applied to the quantum electromagnetic Casimir interaction between plates with roughness \cite{PhysRevLett.87.260402} and between deformed plates \cite{Emig03}.  Finally, the path integral approach was connected to scattering theory in Ref. \cite{Buescher05}.

Because this approach relies on the fundamental formalism of quantum field theory, it has a wide range of applications to problems involving quantum (and, as we will see below, thermal) fluctuations. To our knowledge, it was first developed in precise detail for Casimir calculations in Ref.\ \cite{Langbein1974}.  A typical result in this approach takes the form \cite{Emig06,PhysRevLett.97.160401,spheres}
\begin{equation}
{\cal E}=\frac{\hbar c}{2\pi} \int_0^\infty d\kappa 
\log \det (1 - G_{21} T_1 G_{12} T_2) \,,
\label{eqn:TGTG}
\end{equation}
where the $T_i$ are the scattering $T$-matrices \cite{Waterman65,Waterman71} 
for each object and $G_{12}$ and $G_{21}$ are the free Green's functions, which propagate fluctuations in the scattering basis of one object to the scattering basis of the other.  Here we have considered just two objects, but this approach has a natural extension to a larger numbers of objects \cite{Emig08-1,scalar,PhysRevD.80.085021,PhysRevD.83.045004}.

By making use of the identity
\begin{equation}
\log \det (1-A) = \tr \log (1-A) = - \Tr \sum_{j=0}^\infty \frac{A^j}{j}
\end{equation}
for $A = G_{21} T_1 G_{12} T_2$, we can build an intuitive picture of the physics that this formula captures.  The $T$-matrix $T_1$ describes a quantum fluctuation on object 1.  The Green's function $G_{12}$ then propagates this fluctuation from object 1 to object 2, and the $T$-matrix $T_2$ describes the fluctuation that results from its interaction with object 2, and finally the Green's function $G_{21}$ propagates this result back to object 1.  The sum captures all possible numbers of round-trip reflections, while the trace sums over all possible fluctuation modes with a particular frequency, and the integral sums over all frequencies, here having been Wick rotated to imaginary wave number $i\kappa = k = \omega /c$.

To demonstrate the practical application of the formula in Eq.~\ref{eqn:TGTG}, we consider the simple case of two isotropic particles (such as atoms or nano-particles), which are small compared to their separation $d$. We assume that the particles have only electric dipole polarizabilities $\alpha_1$ and $\alpha_2$, respectively, which are independent of frequency. (The latter is justified in the retarded limit where $d$ is much larger than the speed of light divided by a characteristic frequency below which $\alpha_j$ become static.) Then the Green's functions and the $T$-matrices in Eq.~(\ref{eqn:TGTG}) can be expressed in terms of spherical dipole waves, yielding $3\cross 3$ dimensional matrices. The $T$-matrices are diagonal, given by
\begin{equation}
T_{j;mm'} = \frac{2}{3} \alpha_j \kappa^3 \delta_{mm'}
\end{equation}
for $j=1,2$ and $m=-1,0,1$. With the dipole expansion of the Green's functions $G_{12}$ and $G_{21}$, one obtains after taking the determinant in Eq.~(\ref{eqn:TGTG}) the energy
\begin{equation}
{\cal E}=-\frac{\hbar c}{\pi d} \frac{\alpha_1}{d^3}\frac{\alpha_2}{d^3} \int_0^\infty du (3+6u+5u^2+2u^3+u^4)e^{-2u}
\end{equation}
where we have set $u=\kappa d$ and expanded the logarithm around one since $\alpha_j \ll d^3$. Integration easily yields 
\begin{equation}
{\cal E}=-\frac{23}{4\pi} \hbar c \frac{\alpha_1\alpha_2}{d^7} \,,
\end{equation}
which is the so-called Casimir-Polder potential.

As an exact expression, Eq.\ (\ref{eqn:TGTG}) can be used to prove general results applicable to arbitrary geometries.  For example, Ref.\ \cite{PhysRevLett.97.160401} applies a Feynman-Hellman argument to show that the force between any configuration of electromagnetic conductors with mirror symmetry is always attractive, a result that will inform our discussion below of Casimir stresses on an individual object.  It can also be used to demonstrate a form of Earnshaw's theorem, ruling out the existence of a stable equilibrium generated by Casimir forces \cite{Rahi_2010}.

For experimental work, spherical, cylindrical, and planar geometries are by far the most common and useful.   As first detailed in Ref.\ \cite{Langbein1974}, one can use a multipole basis to express scattering from each object in its own partial wave basis, and then use expansions of the free Green's function to translate scattering from one object's basis to another
\cite{spheres,scalar,PhysRevD.80.085021,Kenneth08}.  This approach has been adapted to a wide range of calculations for planes \cite{Lambrecht06}, cylinder, cylinder-plane, and cylinder-sphere geometries \cite{Emig06,Rahi08-2,PhysRevD.80.085021,Teo:2012kf}, sphere geometries for both scalar \cite{Bulgac06,Wirzba08,scalar} and electromagnetic \cite{spheres,PhysRevD.80.085021} fluctuations, and, most importantly for experiment, sphere-plane geometries in electromagnetism \cite{Emig08-1,Maia_Neto08,canaguier09}.  

\subsection{Geometries with edges and tips}

Scattering methods are much more tractable for objects with geometries for which the $T$-matrix is diagonal.  One can extend the range of such situations by considering more unusual coordinate systems.  For scalar fluctuations, these include elliptic cylinder \cite{Graham2013}, parabolic cylinder \cite{Graham2010}, and spheroidal geometries \cite{Emig2009}, for which limiting cases are a finite-width strip, half-plane, and disk respectively.  One can also consider scattering in a different variable in order to study a wedge or cone using ordinary cylindrical and spherical coordinates, respectively \cite{Maghrebi6867}.  In these cases one uses contour integration to replace the scattering theory sum over partial waves with a continuous integral.  One can also consider complementary geometries using systems with planar gaps \cite{Kabat10} and by applying the generalization of Babinet's principle \cite{Babinet}, and for scalar theories one can apply worldline methods \cite{Gies_2003}.

For perfect conductors, all of these cases can be extended to electromagnetism except for spheroidal scattering, since any translation-invariant geometry can be decomposed into polarizations obeying Dirichlet and Neumann boundary conditions, and scattering from a cone can also be diagonalized through analytic continuation of the standard techniques of Mie scattering.  The limiting case of the perfectly conducting disk can nonetheless be solved exactly as an analytic calculation in oblate spheroidal coordinates\cite{Meixner,Emig2016}, for which the scattering matrix is no longer diagonal.

The wedge, parabolic cylinder, and elliptic cylinder results can be combined to yield a consistent model of the effects of edges \cite{Graham:2011ta,Maghrebi2011,Blose2015}.  This can be expressed as an approximate form for the Casimir energy per unit length of a perfectly conducting strip parallel to a perfectly conducting plane,
\begin{equation}
\frac{\cal E}{\hbar c L} = -\frac{\pi^2}{720} \frac{2d}{H^3}
+ \frac{2 \beta}{H^2} + \frac{\gamma}{2d H} + \ldots
\label{eqn:expansion}
\end{equation}
Here $H$ is the separation between the strip and the plane, $2d$ is the width of the strip, and $\beta=0.00092$ and $\gamma=-0.0040$ are numerical parameters capturing the effect a single edge and the interaction between two edges, respectively; the result for $\beta$ is consistent with limiting cases of the wedge and parabolic cylinder.  One can also obtain the energy per unit length of a perfectly conducting half-plane perpendicular to a perfectly conducting plane as \cite{Graham2010}
\begin{equation}
\frac{{\cal E}}{\hbar c L} = 
\int_0^\infty \frac{q dq}{4\pi} \log \det
\left(\mathbbm{1}_{\nu \nu'} - (-1)^\nu {\rm k}_{-\nu-\nu'-1}(2 q H)\right)
= -\frac{C_\perp}{H^2},
\end{equation}
where $H$ is the separation distance, ${\rm k}_\ell(u)$ is the Bateman k-function \cite{Bateman}, the determinant is over $\nu, \nu' = 0,1,2,3\ldots$,
and $C_\perp=0.0067415$ is obtained by numerical integration.

\subsection{Casimir stresses}

An interesting extension of Casimir force calculations is to the case of Casimir stresses on a single object.  This problem is harder to define than the case of forces between rigid bodies, because deformation of a single object requires a variation of the physical configuration of an object, rather than just a variation of the position of fixed objects.  As a result, the formally infinite contributions localized on the object no longer cancel, and we must apply renormalization techniques, which in turn require a model of the object's material properties.

Debates about Casimir stresses originate in an early model of the electron as a charged shell for which attractive Casimir forces balance Coulomb repulsion \cite{CasimirShell}.  Here the dynamics of the shell are assumed as a definition of the fundamental interactions.  One finds that specifying an ideal conductor boundary condition leads to a positive energy and hence expansion of the shell \cite{Boyer:1968uf,Milton:1978sf,Milton01}, invalidating the model's original premise.  Of course, modern quantum electrodynamics provides a well-verified description of the electron within the broader framework of renormalized quantum field theory.

This result has often been misinterpreted, however, to imply that a conducting shell will experience a repulsive stress.  This case is very different, because the dynamics of an actual piece of conducting material are determined by the underlying physics of the matter that makes it up, and so cannot be assumed to implement an ideal boundary condition at all frequencies.  (The Casimir energy of an idealized boundary can be of interest for the mathematical ``Weyl problem'' of the relationship between eigenvalue spectra and geometry \cite{Abalo:2012jz,Kolomeisky:2010eg}.)  A repulsive stress is also in conflict with the well-established attractive Casimir force between conductors; in particular, the force between two hemispheres, as in any configuration of conductors with mirror symmetry, is always attractive \cite{PhysRevLett.97.160401}.  Similarly, while a boundary condition calculation yields a repulsive stress on a rectangular solid, the force on a piston is attractive, as shown in two dimensions in Ref.\ \cite{Cavalcanti2004} and three dimensions in Ref.\ \cite{PhysRevLett.95.250402}, where the latter is derived using a geometrical optics approximation \cite{PhysRevLett.92.070402}.

The resolution of this puzzle lies in contributions from within the material, which no longer cancel when the body is not rigid.  Modeling the material as a position-dependent dielectric \cite{GRAHAM2013846} shows that the contribution from within the material is attractive, and larger in magnitude than the repulsive external contribution.  In the limit where the dielectric becomes infinitely strong, the sphere becomes a conductor \cite{Bordag:2008rc}, but this limit does not commute with the limits involved in renormalization:  the theory must be cut off at a high enough frequency that the material is transparent.  Qualitatively similar results have been obtained by considering specific materials rather than a generic dielectric, such as a carbon nanostructure \cite{Barton} or a ``fish-eye'' medium \cite{PhysRevD.84.081701}.  These findings are also in agreement with the work of Deutsch and Candelas \cite{Deutsch:1978sc,Candelas1982241}, who showed that divergences in Casimir stresses arise from surface counterterms \cite{Symanzik:1981wd} that cannot be removed by renormalization, as well as more detailed calculations in models with a fluctuating scalar field \cite{Graham:2002fw,Graham:2002xq,Graham:2003ib}.

A key aspect of these works is that divergences associated with an ideal boundary arise in two different ways:  the ``sharp'' limit of infinitesimal thickness, and the ``strong'' limit of strong potential.  For stresses on a real material, these limits are intertwined:  for example, as a spherical shell expands and increases its surface area, it becomes thinner, leading to a weaker electromagnetic response.  A precise description of this situation requires detailed information about the the material's electromagnetic properties, but one can gain a qualitative understanding by approximating this behavior via renormalization counterterms.  In this process, one introduces a short-distance cutoff, which in a material represents a length scale set by the lattice spacing.  The formally divergent contributions arising for a conducting material at short wavelength are then combined with cutoff-dependent counterterms, which are fixed so that the combined result matches experimental inputs.  For a stress calculation, one must then fix the dependence of these inputs on the geometry.  A reasonable estimate is to assume that the overall strength of the potential, integrated over volume, remains constant, roughly corresponding to holding the total number of charge carriers constant.  In scalar models \cite{Graham:2002fw,Graham:2002xq,Graham:2003ib} one can also adjust the strength of the potential used to model the object's scattering response, while for electromagnetic models it is more appropriate to use a Drude model dielectric with fixed plasma wavelength and conductivity, and instead vary the thickness of the material so that its volume remains constant.

\begin{figure}[htbp]
\centerline{\includegraphics[width=0.5\linewidth]{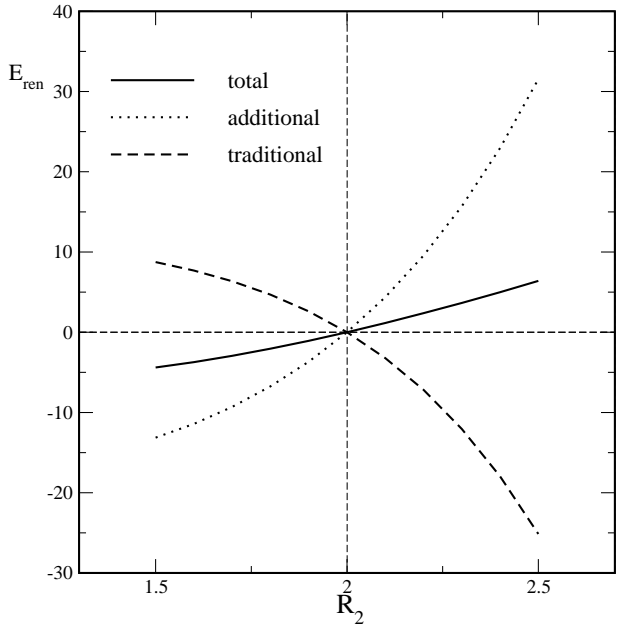}}
\caption{\label{fig_total} Difference in renormalized energy between
shells of radii $R_2$ and $R_1$, as a function of $R_2$ in units where $R_1=2$, for generic values of the Drude plasma wavelength and conductivity.  The attractive effect of the ``additional'' contribution from fluctuations within the material reverses the repulsive ``traditional'' contribution from fluctuations outside the material.}
\label{fig:shell}
\end{figure}

While this result does not precisely model a particular material like the examples listed above, it allows for a generic description of the discrepancy between the apparent repulsive stress on a spherical conducting shell with the generally attractive Casimir force for dieletric materials, including their perfectly conducting limit.  What one finds \cite{GRAHAM2013846} is that the Casimir self-energy decomposes into a sum of two terms:  The ``traditional'' contribution due to the shell's external electromagnetic response and an ``additional'' contribution localized within the thickness of the shell, which consists of a renormalized integral over $r$.  The former reproduces the Boyer result in the conducting limit, but the latter is of comparable magnitude with opposite sign, and renders the full result attractive, as shown in Fig.\ \ref{fig:shell}.

\section{Unified bulk and surface formulation}

\begin{figure}
\includegraphics[scale=0.25]{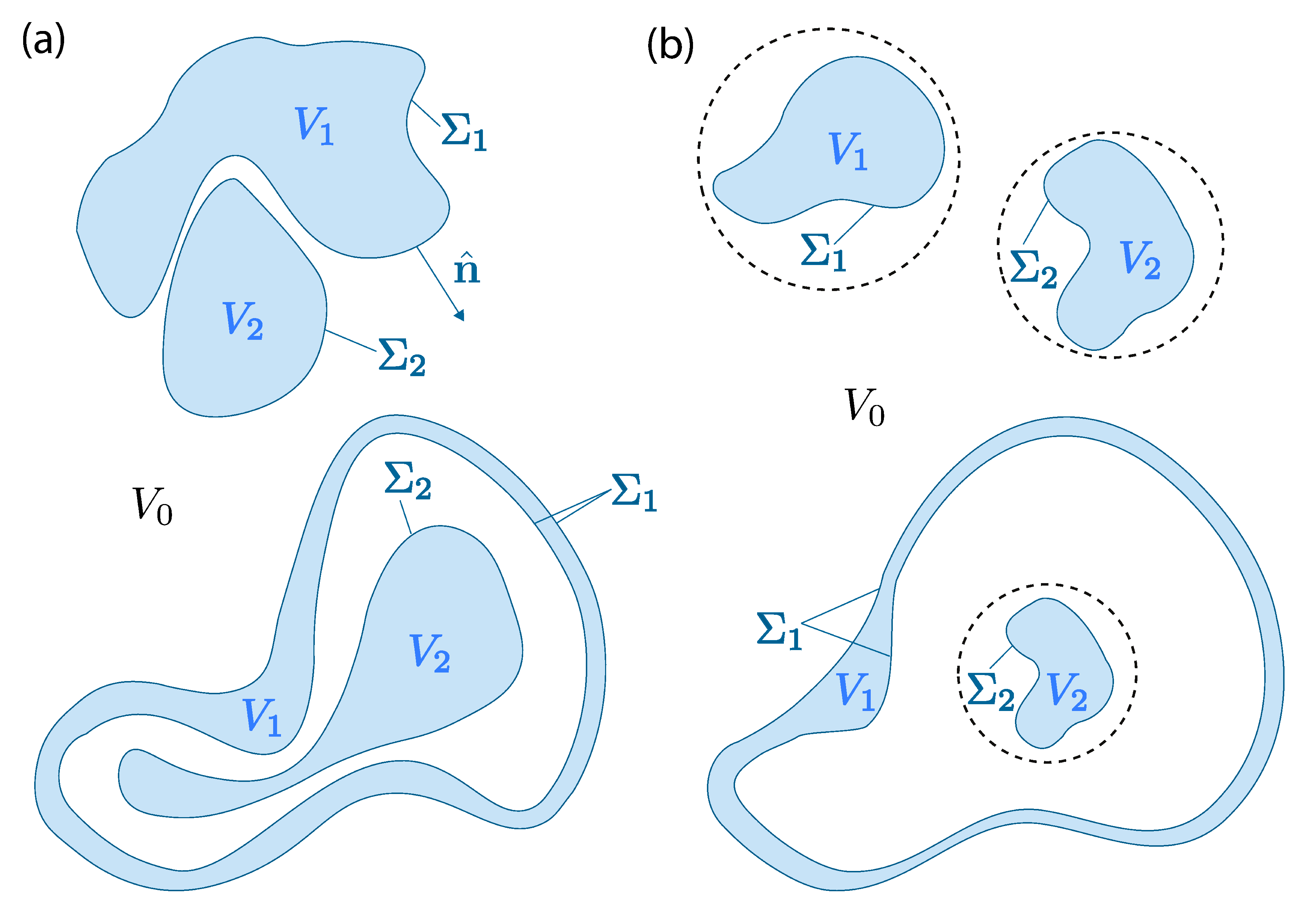}
\caption{Configuration of dielectric bodies: (\textbf{a}) general shapes and positions that can be studied with the approaches reviewed in this section, (\textbf{b}) non-penetrating configurations that can be studied within the scattering approach.\label{fig:configuration}}
\end{figure}  

While the scattering theory method offers an elegant way of expressing Casimir interaction energies that can be used both to obtain analytic results and to carry out precise calculations in semi-analytic form, it also suffers from significant limitations.  For objects without a high degree of symmetry, the $T$-matrix is off-diagonal and therefore more difficult to calculate.  At the same time, as the separation between the objects decreases, higher frequencies and partial waves become important as the calculation becomes dominated by contributions from waves near the point of closest approach \cite{Schaden:1998zz}, so the numerical cutoffs in both the trace over modes and the integral over wave number must be extended to larger and larger values, putting more stress on the numerical algorithms.  Taking this case to a further extreme, one can imagine a situation where objects interpenetrate within the relevant scattering basis --- for example, if scattering from object 1 is computed in a spherical basis and object 2 does not sit entirely outside a sphere enclosing object 1, as shown in Fig.~\ref{fig:configuration}(a) --- then the expansion fails entirely. From an experimental point of view, however, these kinds of short-distance configurations can be the most appealing, since they represent situations where the force is strongest.

Surface methods \cite{Reid:2009aa} can provide an alternative approach that overcomes these limitations.  Originally, the surface approach has been introduced in the literature as a method for a purely numerical computation of Casimir interactions \cite{Johnson_2013}.  There are two different methods to implement the computation of Casimir forces from fluctuating currents. One can either integrate the Maxwell stress tensor over a closed surface enclosing the body, directly yielding  the Casimir force, or integrate over all electromagnetic gauge field fluctuations in a path integral, yielding the Casimir free energy. We shall review  both approaches here. 

Compared to scattering theory-based approaches, the surface formulation has the advantage that it does not require the use of eigenfunctions of the vector wave equation, which are specific to the shapes of the bodies. Hence, this approach is applicable to general geometries and shapes, including interpenetrating structures.  For example, the power of the surface approach has been demonstrated by numerical implementations in Ref.\ \cite{Johnson_2013}, where it was used to compute the Casimir force in complicated geometries.

Recent work has demonstrated the equivalence of a surface current approach and the bulk scattering methods \cite{Bimonte_2021}.  This result can be considered as the Casimir analog of the Huygens equivalence principle \cite{Harrington}, which shows that the electromagnetic effects of an arbitrary current distribution contained within a closed surface can be exactly reproduced by an equivalent current distribution located on the surface. 
The main recent advancements were (i) a new, compact and elegant derivation of the Casimir force from the Maxwell stress tensor within both a $T$-operator approach and a surface operator approach, (ii) a new surface formula for the Casimir free energy expressed in terms of a surface operator, and (iii) a new path integral-based derivation of a Lagrangian and Hamiltonian formulation for the Casimir free energy.  In the following we shall review the main concepts and results of the stress tensor and path integral formulations of the surface current based approach.

The geometries and shapes to which the approaches can be applied are shown in Fig.~\ref{fig:configuration}a. For comparison, in  Fig.~\ref{fig:configuration}b, we display non-penetrating bodies to which scattering theory-based approaches are limited.
In general, we assume a configuration composed of $N$ bodies with dielectric functions $\epsilon_r(\omega)$ and magnetic permeabilities $\mu_r(\omega)$, $r=1,\ldots,N$. The bodies occupy the volumes $V_r$ with surfaces $\Sigma_r$ and outward pointing surface normal vectors $\hat\bfn_r$. The space with volume $V_0$ in between the bodies is filled by matter with dielectric function $\epsilon_0(\omega)$ and magnetic permeability $\mu_0(\omega)$.

\subsection{Stress Tensor Approach}

The common starting point of the bulk and of the surface approaches to computing Casimir forces between two or more  bodies is the physical picture that the vacuum surrounding the bodies is filled with quantum and thermal fluctuations of the EM field.  If the bodies are in thermal equilibrium at temperature $T$ with the environment, the correlators of the EM field can be derived from linear-response theory \cite{agarwal}:
\begin{eqnarray}
\langle {\hat E}_{i}({\bf x},t) {\hat E}_{j}({\bf x'},t') \rangle_{\rm sym} &=& \hbar \int_{-\infty}^{\infty}\!\! \frac{d \omega}{2 \pi} \coth\left(\frac{\hbar \omega}{2 k_B T}\right) {\rm Im}[{\cal G}^{(EE)}_{ij}({\bf x},{\bf x'},\omega)] \;e^{-i \omega(t-t')} \;, \nonumber\\
\langle {\hat H}_{i}({\bf x},t) {\hat H}_{j}({\bf x'},t') \rangle_{\rm sym} &=& \hbar \int_{-\infty}^{\infty}\!\! \frac{d \omega}{2 \pi} \coth\left(\frac{\hbar \omega}{2 k_B T}\right) {\rm Im}[{\cal G}^{(HH)}_{ij}({\bf x},{\bf x'},\omega)] \;e^{-i \omega(t-t')} \;, \nonumber \\
\langle {\hat E}_{i}({\bf x},t) {\hat H}_{j}({\bf x'},t') \rangle_{\rm sym} &=& \hbar \int_{-\infty}^{\infty}\!\! \frac{d \omega}{2 \pi} \coth\left(\frac{\hbar \omega}{2 k_B T}\right)\left\{- {\rm i} \,{\rm Re}[{\cal G}^{(EH)}_{ij}({\bf x},{\bf x'},\omega)] \right\} e^{-i \omega(t-t')} \;,
\label{corr}
\end{eqnarray}   
where the subscript sym on the average symbols denotes the symmetrized products of field operators, and the ${\cal G}_{ij}^{(\alpha \beta)}({\bf r},{\bf r'},\omega)$ with $\alpha,\beta= E,H$ are the Fourier transforms of the classical Green's functions for the system of bodies.  The latter Green's functions are obtained by solving the {\it macroscopic} Maxwell equations, subjected to the appropriate boundary conditions on the surfaces of the bodies.  We underline that Eqs.~(\ref{corr}) include both {\it zero-point} (i.e. quantum) and {\it thermal} fluctuation of the em field.  The validity of Eqs.~(\ref{corr}) is 
subjected to the condition that the length scales of the relevant fluctuations should be large compared to atomic distances, in  such a way that the EM material properties of the bodies can be described by their {\it macroscopic} response functions.  We shall assume for simplicity that the bodies are made out of homogeneous and isotropic magneto-dielectric materials, characterized by the respective frequency-dependent electric and magnetic permittivities $\epsilon(\omega)$ and $\mu(\omega)$. 

The mechanical effects of the fluctuating EM field on the bodies are determined by the expectation value $\langle T_{ij} \rangle$ of the Maxwell stress tensor
\begin{equation}
\langle T_{ij}({\bf x})\rangle=\frac{1}{4 \pi} \left\{ \langle E_i({\bf x}) E_j({\bf x}) \rangle +\langle H_i({\bf x}) H_j({\bf x}) \rangle -\frac{1}{2} \delta_{ij}\left[ \langle E_k({\bf x}) E_k({\bf x}) \rangle + \langle H_k({\bf x}) H_k({\bf x}) \rangle \right]\right\}\;,\label{stress}
\end{equation} 
where the correlators are evaluated for equal times $t=t'$. To gain further insight, it is convenient to  note that at points ${\bf x}, {\bf x}'$ both lying in the vacuum medium, the Green's functions ${\cal G}^{(\alpha \beta)}_{ij}({\bf x},{\bf x}';\omega)$ can be decomposed as follows:
\begin{equation}
{\cal G}^{(\alpha \beta)}_{ij}({\bf x},{\bf x}';\omega)={\cal G}^{(\alpha \beta;0)}_{ij}({\bf x}-{\bf x}';\omega)+{\Gamma}^{(\alpha \beta)}_{ij}({\bf x},{\bf x}';\omega)\;,\label{splitGr}
\end{equation}
where ${\cal G}^{(\alpha \beta;0)}_{ij}({\bf x}-{\bf x}';\omega)$ is the Green's function of free space,  while ${\Gamma}^{(\alpha \beta)}_{ij}({\bf x},{\bf x}';\omega)$  describes the em field {\it scattered} by the bodies. When the above decomposition  is used  in Eq.~(\ref{stress}), one finds that the expectation value of the stress tensor is accordingly decomposed as
\begin{equation}
\langle T_{ij}({\bf x})\rangle= \langle T^{(0)}_{ij}({\bf x})\rangle + {\Theta}_{ij}({\bf x})\;, 
\end{equation} 
where $ \langle T^{(0)}_{ij}({\bf x})\rangle$ is the free-space contribution, while ${\Theta}_{ij}({\bf x})$  is the scattering contribution.  According to Eqs.~(\ref{corr}), $\langle T^{(0)}_{ij}({\bf x})\rangle$ and ${\Theta}_{ij}({\bf x})$ involve frequency integrals of the imaginary parts of the respective Green's functions ${\cal G}^{(\alpha \beta;0)}_{ij}({\bf x},{\bf x}';\omega)$ and ${\Gamma}^{(\alpha \beta)}_{ij}({\bf x},{\bf x}';\omega)$.  Let us consider $\langle T^{(0)}_{ij}({\bf x})\rangle$ first.  Since
\begin{equation}
\lim_{{\bf x} \rightarrow {\bf x}'}{\rm Im}[{\bf \cal G}^{(EE;0)}_{ij}({\bf x},{\bf x}',\omega)]=\lim_{{\bf x} \rightarrow {\bf x}'}{\rm Im}[{\bf \cal G}^{(HH;0)}_{ij}({\bf x},{\bf x}',\omega)]=\frac{2 \, \omega^3}{3 \, c^3} \delta_{ij}\;.\label{freespaceim}
\end{equation}
it is clear that $\langle T^{(0)}_{ij}({\bf x})\rangle$ is expressed by a formally divergent frequency integral.  One notes, however, that according to Eq.~(\ref{freespaceim}),  $\langle T^{(0)}_{ij}({\bf x})\rangle$ represents a {\it homogeneous and isotropic} pressure acting the surfaces of the bodies, which is independent of the material properties of the bodies.   Since  this pressure cannot give rise to an overall force on the body, just a stress as discussed above, we can neglect it here.  Let us turn now to the scattering contribution ${\Theta}_{ij}({\bf x})$.  By performing a Wick rotation to the imaginary frequency axis, one finds that ${\Theta}_{ij}({\bf x})$  has the expression
\begin{eqnarray}
{\Theta}_{ij} ({\bf x}) &=& \frac{k_B T}{2 \pi} \left.\sum_{n=0}^{\infty}\right.\!'     \left[  {\Gamma}^{(EE)}_{ij}({\bf x},{\bf x}; {\rm i}\, \xi_n)
+ {\Gamma}^{(HH)}_{ij}({\bf x},{\bf x}; {\rm i}\, \xi_n) \right. \cr
&& \left. - \frac{1}{2} \delta_{ij} \left({\Gamma}^{(EE)}_{kk}({\bf x},{\bf x}; {\rm i}\, \xi_n)+{\Gamma}^{(HH)}_{kk}({\bf x},{\bf x}; {\rm i}\, \xi_n)\right)   \right]\;. \label{stressGamma}
\end{eqnarray}
where $\xi_n= 2 \pi n k_B T/\hbar$ are the Matsubara imaginary frequencies, and the prime in the summations means that the $n=0$ term is taken with a weight of one-half. The sum on the right-hand side of the above equation is {\it finite}, because the scattering parts of the Green's functions  ${\Gamma}^{(\alpha \beta)}_{ij}({\bf x},{\bf x}';\omega)$ remain finite in the coincidence limit  ${\bf x}' \rightarrow {\bf x}$.

The Casimir force on body $r$ can be now obtained by integrating ${\Theta}_{ij}({\bf x})$ on any surface $S_r$ drawn in the vacuum that surrounds that body and excludes all other bodies, yielding
\begin{equation}
F_i^{(r)} =\oint_{{S}_r} d^2 \sigma  \,\hat { n}_j({\bf x})\;  {\Theta}_{ji} ({\bf x}) \;.\label{force0}
\end{equation} 
The surface integral on the right-hand side of this equation can be recast in a remarkably simple form, by taking advantage of the peculiar structure of the scattering Green's functions ${\Gamma}^{(\alpha \beta)}_{ij}({\bf x},{\bf x}')$ (for brevity, from now on we shall not display the dependence of the Green's functions on the Matsubara frequencies $\xi_n$).   This structure becomes manifest once ${\Gamma}^{(\alpha \beta)}_{ij}({\bf x},{\bf x}')$ is expressed in terms of the so-called $T$-operator \cite{Waterman65,Waterman71}. We recall that the $T$-operator $T_{kl}^{(\rho \sigma)}({\bf y},{\bf y}')$ is defined to give the currents induced at the point ${\bf y}$ in the interior of the bodies  by an external em field at point ${\bf y}'$,  Using the $T$-operator, the scattering parts of the Green's functions can be expressed as  
\begin{equation}
{\Gamma}^{(\alpha \beta)}_{ij}({\bf x},{\bf x}' )= \sum_{r,r'=1}^{N} \int_{V_r}  d^3 {\bf y} \int_{V_{r'}} d^3 {\bf y}'  {\cal G}^{(\alpha \rho;0)}_{ik}({\bf x}-{\bf y} )    T_{kl}^{(\rho \sigma)}({\bf y},{\bf y}')   {\cal G}_{lj}^{(\sigma \beta;0)}({\bf y}'-{\bf x}')\;.\label{repGreen1}
\end{equation}  
Despite its simple meaning, the $T$-operator is difficult to compute in general, and its expression is known only for bodies with sufficiently symmetric geometries, mostly in cases where the $T$-operator is diagonal. A more powerful representation of the scattering Green's functions can be  derived from the {\it equivalence principle} of classical electromagnetic theory \cite{Harrington}.  According to this principle, the induced currents existing in the interior of the bodies can be replaced by  {\it fictitious} surface currents,  which  produce  the {\it same} scattered field as  the induced  currents.  By following this equivalence principle, one arrives at an  alternative representation of  the  scattering Green's functions consisting of a double  {\it surface} integral extended onto the surfaces of the bodies \cite{Johnson_2013,Bimonte_2021}: 
\begin{eqnarray}
{\Gamma}^{(\alpha \beta)}_{ij}({\bf x},{\bf x}' )&=&- \sum_{r,r'=1}^{N} \int_{V_r}  d^3  {\bf y} \int_{V_{r'}} d^3 {\bf y}'  \,\delta(F_r ({\bf y}))\, \delta(F_{r'}({\bf y}'))\;  \nonumber \\
&\times&{\cal G}^{(\alpha \rho;0)}_{ik}({\bf x}- {\bf y}) )    \left(M^{-1}\right)_{kl}^{(\rho \sigma)}({\bf y},{\bf y}')   {\cal G}_{lj}^{(\sigma \beta;0)}({\bf y}'-{\bf x}')\;,\label{repGreen2}
\end{eqnarray}
where $\delta(x)$ is the Dirac delta function and $F_r({\bf y})=0$ is the equation of the surface $\Sigma_r$ of the $r$-th body.
The {\it surface} operator $M_{ij}^{(\alpha \beta)}({\bf y},{\bf y}')$, whose inverse appears in the integral on the right-hand side, has a remarkably simple expression \cite{Johnson_2013,Bimonte_2021}
\begin{equation}
M_{ij}^{(\alpha \beta)}({\bf y},{\bf y}') = \left\{ \begin{array}{ll}
\!  \Pi^{(r)}_{ik} ({\bf y}) \left[{\cal G}^{(\alpha \beta; r)}_{kl}({\bf y}-{\bf y}')  + {\cal G}^{(\alpha \beta; 0)}_{kl}({\bf y}-{\bf y}') \right] \Pi^{(r)}_{lj}({\bf y}')   & {\rm if} \;\; {\bf y}, {\bf y}'  \in \Sigma_r \\
 \Pi_{ik}^{(r)} ({\bf y}) \; {\cal G}^{(\alpha \beta; 0)}_{kl}({\bf y}-{\bf y}') \;\Pi_{lj}^{(s)}({\bf y}')  & {\rm if} \;\, {\bf y} \in \Sigma_r , {\bf y}'  \in \Sigma_s , \\ & \;\;\;\; r \neq s
\end{array}\right. \, .\label{asF}
\end{equation}
where $\Pi^{(r)}_{ik} ({\bf y})$ is the projector onto the plane tangent at $\Sigma_r$ at ${\bf y}$, and ${\cal G}^{(\alpha \beta; r)}_{ij}({\bf y}-{\bf y}')$ are the Green's functions for an {\it infinite} homogeneous and isotropic magneto-dielectric  medium identical to that of body $r$.  By inspection of Eqs.\ \ref{repGreen1} and \ref{repGreen2}, it is apparent that both representations of the scattering Green's functions have a common ${\hat G}^{(0)} {\hat {\cal K}}\, {\hat G}^{(0)}$ structure, where the ${\hat {\cal K}}$ operator coincides either with Waterman's $T$-operator, or with (minus) the inverse of the surface operator ${\hat M}$.  Importantly, both the $T$-operator and the surface operator ${\hat M}$ satisfy  the reciprocity relations \cite{Harrington} enjoyed by the Green's functions.  In Ref.~\cite{Bimonte_2021} it is shown that whenever the scattering Green's function has the ${\hat G}^{(0)} {\hat {\cal K}}\, {\hat G}^{(0)}$ structure, with a ${\hat {\cal K}}$ satisfying reciprocity,  the force formula in Eq.~(\ref{force0}) can be recast in the remarkably simple form
\begin{equation}
{\bf F}^{(r)}=k_B T  \left.\sum_{n=0}^{\infty}\right.\!\!'  \,{\rm Tr} \left[ {\hat {\cal K}}({\rm i}\, \xi_n) \frac{\partial}{\partial {\bf x}_r} \hat{\cal G}^{(0)} ({\rm i}\, \xi_n)\right]\;,\label{force}
\end{equation}
where the trace operation ${\rm Tr}$  denotes an integral over the volumes occupied by the bodies and a sum over both the spatial indices $i,j$ and internal indices $\alpha, \beta$, while the symbol $ {\partial}/{\partial {\bf x}_r}$ denotes a derivative with respect to a rigid translation of the $r$-th body.  When ${\hat {\cal K}}$ is identified with the $T$-operator, Eq.~(\ref{force})  reproduces the ``TGTG'' formula derived in Ref.\ \cite{PhysRevLett.97.160401}.  If ${\hat {\cal K}}$ is instead identified with (minus) the surface operator $(\hat M)^{-1}$, then Eq.~(\ref{force}) reproduces the surface force formula derived in Ref.~\cite{Johnson_2013}. We thus see that Eq.~(\ref{force}) encompasses in a single compact formula the bulk and the surface formulations of the Casimir force.

Starting from Eq.~(\ref{force}), it is possible to compute the Casimir free-energy of the system of bodies.  Direct integration of Eq. (\ref{force}) leads to a formally divergent result ${\cal F}_{\rm bare}$.  However, a finite expression is easily recovered by subtracting from  ${\cal F}_{\rm bare}$ the self-energies of the individual bodies. The details of the subtraction procedure differ slightly within the bulk and surface formulations, but the  final formula for the renormalized Casimir free energy ${\cal F}$  has an identical form in both approaches \cite{Bimonte_2021}. In the simple case of two bodies, it reads
\begin{equation}
{\cal F}=k_B T \left. \sum_{n=0}^{\infty}\right.\!'  {\rm Tr}\,\log \left[1- \hat{\cal{K}}_1 \, \hat{\cal G}^{(0)} \hat{\cal{K}}_2 \hat{\cal G} ^{(0)} \right]\;.\label{renensurfgen}
\end{equation}
Within the bulk approach, the operator $\hat{{\cal K}}_r$ coincides with the $T$-operator $T_{ij}^{(\alpha \beta;r)}({\bf y},{\bf y}')$  of body $r$ in isolation, while in the surface approach $\hat{{\cal K}}_r$ coincides with (minus) the inverse of the surface operator  $M_{ij}^{(\alpha \beta;r)}({\bf y},{\bf y}')$ defined by the first line of Eq.\ \ref{asF}.

We underline that Eqs.~(\ref{force}) and (\ref{renensurfgen}) are valid for any shapes and relative dispositions of the bodies. In particular, both Equations hold for a system of {\it interleaved} bodies, possibly nested one inside the other.   A case deserving special consideration is that of  a ``separable'' configuration  of two bodies, namely a configuration in which the bodies can be separated from one another by a plane drawn in the vacuum between them.  Three typical examples of separable configurations are those of a sphere opposite a plate, a system of two spheres, or a sphere and a cylinder, which represent  the configurations adopted in the vast majority of experiments. In \cite{Bimonte_2021} it is shown that in a separable configuration, the general formula Eq.\ \ref{renensurfgen} reproduces the famous scattering formula \cite{spheres,Kenneth08,Lambrecht06,PhysRevD.80.085021}
in which the operators $\hat{{\cal K}}_r$ are replaced by the 
the {\it scattering matrices} of the bodies and $\hat{\cal G} ^{(0)}$ is replaced by the so-called translation matrices.

\subsection{Path Integral Approach}

As in the previous subsection, we consider again $N$ dielectric bodies with properties described above.  In the Euclidean path integral quantization of the electromagnetic field, the Casimir free energy at finite temperature $T$ can be obtained as
\begin{equation}
\label{PI_free_energy}
\cF = - k_B T \left. \sum_{n=0}^{\infty}\right.\!'
\log\frac{\cZ(\xi_n)}{\cZ_{\infty}(\xi_n)} \, ,
\end{equation}
where again the sum runs over the Matsubara frequencies $\xi_n=2\pi n k_B T/\hbar$, with a weight of $1/2$ for $n=0$. The partition function $\cZ$ is given by a path integral that we shall derive now. The partition function $\cZ_{\infty}$ describes the configuration of infinitely separated bodies and subtracts the self-energies of the bodies from the bare free energy.
 
We express the action of the electromagnetic field in the absence of free sources in terms of the gauge field $\bfA$. We chose the transverse gauge with $A_0=0$. The functional integral will then run over $\bfA$ only. The electric field is given by $\bfE = i k \bfA \to -\k \bfA$ and the magnetic field by $\bfB = \nabla \times \bfA$. Then the action  in terms of the induced sources at fixed frequency $\k$ is given by
\begin{eqnarray}
\label{eq:EM_hatS}
  \hat S[\bfA] &=& -\frac{1}{2} \int_{\mathbb{R}^3} d^3 {\bf x}\, \left[
                        \bfA^2 \epsilon_\bfx \k^2 + \frac{1}{\mu_\bfx} (\nabla \times \bfA)^2\right]
- \k \sum_{r=1}^N \int_{V_r} d^3 {\bf x} \,\bfA \cdot \bfP_r \, ,
\end{eqnarray}
for fluctuations $\bfA$ of the gauge field, and induced bulk
currents $\bfP_r$ inside the objects. The inverse of the kernel of the quadratic part of this action is given by the Green's tensor which for spatially constant $\epsilon_r$ and $\mu_r$ of body $r$ is given by
\begin{equation}
  \label{eq:G_dyadic}
  {\stackrel{\leftrightarrow}{\cal G}}^{(AA;r)}(\bfx,\bfx') = \mu_r\left( {\bf 1}
    -\frac{1}{\epsilon_r \mu_r\k^2} \nabla\otimes\nabla \right)
  \frac{e^{-\sqrt{\epsilon_r \mu_r}\k |\bfx-\bfx'|}}{
    |\bfx-\bfx'|} \, ,
\end{equation}
From the relation between the gauge field $\bfA$ and the electric field $\bfE$ the relation $-\k^2{\stackrel{\leftrightarrow}{\cal G}}^{(AA;r)}(\bfx,\bfx')={\stackrel{\leftrightarrow}{\cal G}}^{(EE;r)}(\bfx,\bfx')$ follows, which allows us to compare the results below to those of the stress-tensor-based derivation. Next, we define the classical solutions $\cbA_r$ of the vector wave equation in each region $V_r$, obeying $\nabla \times \nabla \times \cbA_r + \epsilon_r \mu_r \k^2 \cbA_r  = - \k \mu_r \bfP_r $.  Then the source terms of Eq.~(\ref{eq:EM_hatS}) can be written as an integral over the surface of body,
\be
  \label{eq:EM_source_term_1}
  -\k \int_{V_r} d^3 {\bf x} \bfA \cdot \bfP_r =
                                                             \frac{1}{\mu_r} \int_{\Sigma_r} d^3 {\bf x}  \left[ \bfA_-\cdot
      (\bfn_r \times (\nabla \times \cbA_r)) + (\nabla \times \bfA)_-  \cdot (\bfn_r \times \cbA_r )\right] \, .
\ee
The values of the gauge field $\bfA$
and its curl $\nabla\times\bfA$ appearing in this expression are those when the surface is approached from the inside, denoted by  $\bfA_-$ and $(\nabla\times\bfA)_-$. It is important to note that in the above surface integral, $\bfA$ and $\nabla\times\bfA$ multiply vectors that are tangential to the surface, and hence only the tangential components of $\bfA$ and $\nabla\times\bfA$ contribute to the integral. Hence, we can use the continuity conditions of the tangential components of $\bfE$ and $\bfH$,
\begin{equation}
  \bfn_r \times \bfE_- =  \bfn_r \times \bfE_+\, , \quad
  \frac{1}{\mu_r}\bfn_r \times (\nabla \times \bfE)_- =
  \frac{1}{\mu_0}\bfn_r \times  (\nabla \times \bfE)_+ \, ,
\end{equation}
to write the source terms also as
\be
  \label{eq:EM_source_term_2}
  -\k \int_{V_r} d^3 {\bf x} \bfA \cdot \bfP_r =
   \int_{\Sigma_r} d^3 {\bf x}  \left[  \frac{1}{\mu_r}\bfA_+ \cdot
      (\bfn_r \times (\nabla \times \cbA_r)) +  \frac{1}{\mu_0} (\nabla \times \bfA)_+  \cdot (\bfn_r
      \times \cbA_r )  \right] \, , 
\ee
where the values of the gauge field $\bfA$
and its curl $\nabla\times\bfA$ appearing in this expression are now those when the surface is approached from
the outside, denoted by  $\bfA_+$
and $(\nabla\times\bfA)_+$.

Now we shall see the advantage of having expressed the source integrals in terms of the values of $\bfA$ and $\nabla\times\bfA$ when the surfaces are approached from either the outside or the inside of the objects. In the region $V_0$, the field $\bfA\equiv\bfA_0$ is fully determined by its values on the surfaces $\Sigma_r$ and  $\epsilon_0$, $\mu_0$, which are constant across $V_0$. When integrating out $\bfA_0$, one computes the two-point correlation function of $\bfA_+$ and $(\nabla\times\bfA)_+$ {\it on} the surfaces $\Sigma_r$, and hence the behavior of $\bfA_0$ inside the regions $V_r$ with $r>0$ is irrelevant. Following the same arguments for $\bfA\equiv\bfA_r$ inside the objects, the behavior of $\bfA_r$ outside of region $V_r$ is irrelevant for computing the correlations of $\bfA_-$ and $(\nabla\times\bfA)_-$ {\it on} the surfaces $\Sigma_r$. 
Hence, we can replace in the action the spatially dependent $\epsilon_\bfx$ by $\epsilon_0$ when the coupling of $\bfA_0$ to the surface fields $\cbA_r$ is represented by  Eq.~(\ref{eq:EM_source_term_2}), and similarly replace $\epsilon_\bfx$ by $\epsilon_r$ when the coupling of $\bfA_r$ to the surface fields $\cbA_r$ is represented by Eq.~(\ref{eq:EM_source_term_1}).

That this is justified can also be  understood as follows. The field $\bfA_0$ in region $V_0$ can be expanded in a basis of functions that obey the wave equation with $\epsilon_0$. The same can be done for $\bfA_r$ in the interior of each object, i.e., $\bfA_r$ can be expanded in a basis of functions that obey the wave equation with $\epsilon_r$ in $V_r$. For each given set of expansion coefficients in $V_0$ there are corresponding coefficients within each region $V_r$ that are determined by the continuity conditions at the surfaces $\Sigma_r$. The functional integral over $\bfA$ then corresponds to integrating over consistent sets of expansion coefficients that are related by the continuity conditions. The two-point correlations of $\bfA_+$ and  $(\nabla\times\bfA)_+$ on the surfaces $\Sigma_r$ are then fully determined by the integral over the expansion coefficients of $\bfA_0$ in $V_0$ only, and the interior expansion coefficients play no role. Equivalently, the two-point correlations of $\bfA_-$ and  $(\nabla\times\bfA)_-$ on the surfaces $\Sigma_r$ are then fully determined by the integral over the expansion coefficients of $\bfA_r$ in $V_r$ only, and now the exterior expansion coefficients are irrelevant. Hence, in the functional integral, the integration of $\bfA$ can be replaced by $N+1$ integrations over the fields $\bfA_r$, $r=0,\ldots,N$, where each $\bfA_r$ is allowed to extend over unbounded space with the action for a free field in a homogeneous space with $\epsilon_r$, $\mu_r$.
The  multiple counting of degrees of freedom that results from $N+1$ functional integrations poses no problem since the (formally infinite) factor in the partition function cancels when the Casimir energy is computed from Eq.~(\ref{PI_free_energy}).

With this representation, we can write the partition function as a functional integral over the fluctuations $\bfA_r$, separately in each region $V_r$, and the surface fields $\cbA_r$ on body $r$, with the action
\begin{eqnarray}
\hat S[\{\bfA_r\},\{\cbA_r\}]&=&-\frac{1}{2}
                                           \sum_{r=0}^N\int_{\mathbb{R}^3}
                                           d^3 {\bf x}\left[  \bfA_r^2 \epsilon_r \k^2 + \frac{1}{\mu_r}
                        (\nabla \times \bfA_r)^2 \right]
  \\ \nonumber
                                       &+&
                                           \sum_{r=1}^N\int_{\Sigma_r} d^3 {\bf x} \, \left[  \frac{1}{\mu_r}\bfA_0
      (\bfn_r \times (\nabla \times \cbA_r)) + \frac{1}{\mu_0} (\nabla \times \bfA_0) (\bfn_r
                                                             \times
                                                             \cbA_r
                                                             )
                                                             \right]\\ \nonumber
&+& \sum_{r=1}^N\int_{\Sigma_r} d^3 {\bf x} \,\left[ \frac{1}{\mu_r}\bfA_r
      (\bfn_r \times (\nabla \times \cbA_r)) + \frac{1}{\mu_r} (\nabla \times \bfA_r) (\bfn_r
                                                             \times
                                                             \cbA_r
                                                             ) \right] , .
\end{eqnarray}
Now, the fluctuations $\bfA_r$ can be integrated out easily, noting that the two-point correlation function $\langle \bfA_r(\bfx)
\bfA_{r'}(\bfx')\rangle=0$ for all $r$, ${r'}=0,\ldots,N$ with
$r\neq {r'}$. This yields the partition function
\begin{eqnarray}
\label{eq:euclid-z-2}
\cZ(\xi) &=& \prod_{r=1}^N \int \cD\cbA_r
\exp \left[ -\frac{\beta}{2} \left( 
\sum_{r=1}^N\int_{\Sigma_{r}}\!\!\!d^3 {\bf x} \int_{\Sigma_{r}}\!\!\! d^3 {\bf x}' \cbA_r(\bfx) L_{r}(\bfx,\bfx') \cbA_r(\bfx') \right. \right. \nonumber \\
 &+& \left.\left.
\sum_{r,{r'}=1}^N\int_{\Sigma_{r}}\!\!\!d^3 {\bf x} \int_{\Sigma_{{r'}}} \!\!\!d^3 {\bf x}' \cbA_r(\bfx) M_{r{r'}}(\bfx,\bfx') \cbA_{r'}(\bfx')\right)\right]
\end{eqnarray}
with the kernels
\begin{align} 
%[box=\fbox]{align}
%\begin{eqnarray}
\label{eq:kernel_M}
L_{r}(\bfx,\bfx') = &\frac{1}{\mu_r^2} \left[ \,
  \nabla \times \nabla \times {\stackrel{\leftrightarrow}{\cal G}}^{(AA;r)}(\bfx,\bfx') (\bfn_r
  \times \cev{\cdot}\,) (\bfn'_r
  \times \vec{\cdot}\,)\right.\nonumber\\
  &+ \left.\nabla \times
  {\stackrel{\leftrightarrow}{\cal G}}^{(AA;r)}(\bfx,\bfx') (\bfn_r 
  \times (\nabla \times \cev{\cdot}\,)) (\bfn'_r
  \times \vec{\cdot}\,) \right. \nonumber \\
  &+ \left. \nabla \times
  {\stackrel{\leftrightarrow}{\cal G}}^{(AA;r)}(\bfx,\bfx') (\bfn_r
  \times \cev{\cdot}\,) (\bfn'_r 
  \times (\nabla' \times \vec{\cdot}\,)) \right. \nonumber\\
&+  \left. {\stackrel{\leftrightarrow}{\cal G}}^{(AA;r)}(\bfx,\bfx')  (\bfn_r 
  \times (\nabla \times \cev{\cdot}\,))  (\bfn'_r 
  \times (\nabla' \times \vec{\cdot}\,)) \right] \nonumber \\
M_{r{r'}}(\bfx,\bfx') = &\frac{1}{\mu_0^2} \,
 \nabla \times \nabla \times {\stackrel{\leftrightarrow}{\cal G}}^{(AA;0)}(\bfx,\bfx') (\bfn_r
  \times \cev{\cdot}\,) (\bfn'_{r'}
  \times \vec{\cdot}\,) \nonumber\\
  &+
\frac{1}{\mu_0\mu_r} \, 
\nabla \times  {\stackrel{\leftrightarrow}{\cal G}}^{(AA;0)}(\bfx,\bfx') (\bfn_r 
  \times (\nabla \times \cev{\cdot}\,)) (\bfn'_{r'}
  \times \vec{\cdot}\,)\nonumber\\
&+ \frac{1}{\mu_0\mu_{r'}} \, \nabla \times  {\stackrel{\leftrightarrow}{\cal G}}^{(AA;0)}(\bfx,\bfx') (\bfn_r
  \times \cev{\cdot}\,) (\bfn'_{r'} 
  \times (\nabla' \times \vec{\cdot}\,))\nonumber\\
&+ \frac{1}{\mu_r\mu_{r'}}  \,  {\stackrel{\leftrightarrow}{\cal G}}^{(AA;0)}(\bfx,\bfx') (\bfn_r 
  \times (\nabla \times \cev{\cdot}\,))  (\bfn'_{r'} 
  \times (\nabla' \times \vec{\cdot}\,)) \, ,
%\end{eqnarray}
\end{align}
where the arrow over the placeholder $\cdot$ indicates to which side of the kernel it acts. This notation implies that the derivatives are taken before the kernel is evaluated with $\bfx$ and $\bfx'$ on the surfaces $\Sigma_r$, i.e., information about the behavior of basis functions is required in an infinitesimal vicinity of the surfaces.  There is an important simplification of this representation: The bilinear form described by the kernel $L_r$ is degenerate on the space of functions over which the functional integral runs, i.e.,
$\int_{\Sigma_{r}}\!\!\!d^3 {\bf x} \int_{\Sigma_{r}}\!\!\! d^3 {\bf x}'
\cbA_r(\bfx) L_{r}(\bfx,\bfx') \cbA_r(\bfx')=0$ for all 
$\cbA_r(\bfx)$ that are regular
solutions of the vector wave equation 
$\nabla \times \nabla \times \cbA_r + \epsilon_r \mu_r
  \k^2 \cbA_r  =0$ inside region
$V_r$.  This implies that the kernel $L_r$ can be
ignored in the above functional integral over regular waves $\cbA_r$ inside the objects, and the partition function in Eq.\ \ref{PI_free_energy} is given by the functional determinant of the kernel $M_{rr'}$ alone, which can be computed in some basis for regular waves inside the bodies, evaluated at the surfaces only.

The obtained representation of the partition function sums over all configurations of the surface fields $\cbA_r$, and the action depends both on  $\cbA_r$ and the tangential part of its curl, which is functionally dependent on $\cbA_r$. Hence, the situation is similar to classical mechanics where the Lagrangian depends on the trajectory $q(t)$ and its velocity $\dot q(t)$. The Lagrangian path integral runs then over all of path $q(t)$ with $\dot q(t)$ determined by the path automatically. To obtain a representation in terms of a space of functions that are defined strictly on the surfaces $\Sigma_r$ only, 
it would be useful to be able to integrate over $\cbA_r$ and its derivatives {\it independently}. In classical mechanics, this is achieved by Lagrange multipliers that lead to a Legendre transformation of the action to its Hamiltonian form. 
Here the situation is similar. Let us consider the part of the action, $S_r$, which, after functional integration  over $\bfA_r$, generates the kernel $L_r$ which above was shown to vanish. 
The exponential of this part of the action can be written as a functional integral
over two new vector fields $\bfK_r$ and $\bfK'_r$ that are defined
on the surfaces $\Sigma_r$ and are {\it tangential} to the surfaces,
\begin{align}
  \label{eq:EM_rep_kernel_L}
 & \exp(-\beta S_r) \nonumber \\  &=  \cZ_r \oint \cD \bfK_r \cD
                            \bfK'_r
                            \exp \left\{ - \frac{\beta}{2} \frac{1}{\mu_r^2}
                            \int_{\Sigma_r} d^3 {\bf x}
                            \int_{\Sigma_r} d^3 {\bf x}' \left[
                            \bfK_r(\bfx) \cdot \nabla\times\nabla\times
                            {\stackrel{\leftrightarrow}{\cal G}}^{(AA;r)}(\bfx,\bfx') \cdot
                            \bfK_r(\bfx') \right. \right. 
                            \nonumber\\
                            &+ \left.  \left. \bfK_r(\bfx) \cdot
                                \nabla \times {\stackrel{\leftrightarrow}{\cal G}}^{(AA;r)}(\bfx,\bfx') \cdot
                                \bfK'_r(\bfx') +\bfK'_r(\bfx) \cdot
                                \nabla \times {\stackrel{\leftrightarrow}{\cal G}}^{(AA;r)}(\bfx,\bfx')
                               \cdot  \bfK_r(\bfx') 
\right. \right. \nonumber \\                                
                                &+ \left. \left. \bfK'_r(\bfx) \cdot
                                {\stackrel{\leftrightarrow}{\cal G}}^{(AA;r)}(\bfx,\bfx') \cdot
                                \bfK'_r(\bfx')
                            \right] \right.
                            \nonumber\\
  & + \left. \frac{1}{\mu_r} \int_{\Sigma_r} d^3 {\bf x} \,\left[
      \bfA_r \cdot \big(
      (\bfn_r \times (\nabla \times \cbA_r)) - \bfK'_r \big) + (\nabla \times \bfA_r) \cdot \big((\bfn_r
                                                             \times
                                                             \cbA_r
                                                            -\bfK_r) \big) \right] \right\} \, , 
\end{align}
where $\cZ_r$ is some normalization coefficient, and we have used $\oint \cD \bfK_r \cD \bfK'_r$ to indicate that the functional integral extends only over vector fields that are tangential to the surface $\Sigma_r$.   This representation shows that the $\bfA_r$ acts as Lagrange multiplier. Integration over this field removes the imposed constraints between the {dependent} tangential fields $\bfn_r\times\cbA_r$, $\bfn_r \times (\nabla \times \cbA_r)$ by replacing them with the independent tangential surface currents $\bfK_r$ and $\bfK'_r$, respectively.

Substituting Eq.~(\ref{eq:EM_rep_kernel_L}) for each object into the expression for the partition function, integrating out the fields $\bfA_r$ for $r=1,\ldots,N$, constraining the functional integral over $\cbA_r$ to be replaced by the substitutions $\bfn_r \times\cbA_r \to \bfK_r$ and $\bfn_r \times (\nabla \times \cbA_r) \to \bfK'_r$, and finally integrating out $\bfA_0$  yields
\begin{align}
\label{eq:euclid-z-3-EM}
\cZ(\xi) = & \prod_{r=1}^N \oint \cD\underline{\bfK}_r
\exp \left[ -\frac{\beta}{2} \left( 
\sum_{r=1}^N\int_{\Sigma_{r}}\!\!\!d^3 {\bf x} \int_{\Sigma_{r}}\!\!\! d^3 {\bf x}' \, \underline{\bfK}_r(\bfx) \hat L_{r}(\bfx,\bfx') \underline{\bfK}_r(\bfx') \right. \right. \nonumber \\
&+
\left.\left.\sum_{r,{r'}=1}^N\int_{\Sigma_{r}}\!\!\!d^3 {\bf x} \int_{\Sigma_{{r'}}} \!\!\!d^3 {\bf x}' \, \underline{\bfK}_r(\bfx) \hat M_{r{r'}}(\bfx,\bfx') \underline{\bfK}_{r'}(\bfx')\right)\right]\, ,
\end{align}
with the  kernels
\begin{align}
\label{eq:kernel_M_2_EM}
\hat L_{r}(\bfx,\bfx') & = \frac{1}{\mu_r^2}
\begin{pmatrix}
\nabla\times\nabla\times {\stackrel{\leftrightarrow}{\cal G}}^{(AA;r)}(\bfx,\bfx') & \nabla\times {\stackrel{\leftrightarrow}{\cal G}}^{(AA;r)}(\bfx,\bfx') \\\nabla\times {\stackrel{\leftrightarrow}{\cal G}}^{(AA;r)}(\bfx,\bfx') &
{\stackrel{\leftrightarrow}{\cal G}}^{(AA;r)}(\bfx,\bfx')
\end{pmatrix} \\
\hat M_{r{r'}}(\bfx,\bfx') &= 
\begin{pmatrix}
\frac{1}{\mu_0^2} \, \nabla\times\nabla\times {\stackrel{\leftrightarrow}{\cal G}}^{(AA;0)}(\bfx,\bfx') &
\frac{1}{\mu_0 \mu_{r'}} \, \nabla\times {\stackrel{\leftrightarrow}{\cal G}}^{(AA;0)}(\bfx,\bfx') \\
\frac{1}{\mu_0 \mu_r}\, \nabla\times {\stackrel{\leftrightarrow}{\cal G}}^{(AA;0)}(\bfx,\bfx')
&\frac{1}{\mu_r \mu_{r'}} \,  {\stackrel{\leftrightarrow}{\cal G}}^{(AA;0)}(\bfx,\bfx')
\end{pmatrix}  \, .
\end{align}
It should be noted again that the functional integral in Eq.~(\ref{eq:euclid-z-3-EM}) runs over {tangential} vector fields $\bfK_r$, $\bfK'_r$ defined on the surfaces $\Sigma_r$ only.  The kernels $\hat L$ and $\hat M$ can be combined into the joint kernel
$\hat N_{r{r'}}=\hat L_r \delta_{r{r'}} + \hat M_{r{r'}}$. With this kernel, the Casimir free energy is  given by
\begin{equation}
\cF = - k_B T \sum_{n=0}^{\infty}\! \!' \log  \det \left[\hat N(\k_n) \hat N_\infty^{-1}(\k_n)\right] \,\label{hamen} ,
\end{equation}
where the determinant runs over all indices, i.e., $\bfx$, $\bfx'$
located {on} the surfaces $\Sigma_r$, and $r$, ${r'}=1,\ldots,N$. The kernel $\hat N_\infty$ is obtained from the kernel $\hat N$ by taking the distance between all bodies to infinity, i.e, by setting $\hat M_{r{r'}} =0$ for all $r\neq {r'}$. 

The stress tensor approach and the Hamiltonian version of the path integral representation is equivalent to the one derived by Johnson et al.\ as a purely numerical approach using Lagrange multipliers to enforce the boundary conditions in the path integral \cite{Johnson_2013}. Interestingly, the derivation of this representation presented here from a Lagrangian path integral demonstrates the relation of this approach to the scattering approach when the $T$-matrix is defined, as originally by Waterman, by surface integrals of regular solutions of the wave equation over the bodies' surfaces \cite{Waterman65}. This shows the close connection of these approaches, motivating further research in the direction of new semi-analytical methods to compute Casimir forces.

\section{Discussion and Open Problems}

Although the Casimir energy is most often described as a force arising from fluctuations in empty space, as we have seen it can also be profitably re-expressed in terms of surface currents.  One particularly striking open problem for which surface approaches may offer new insights is the contribution of zero-frequency modes to Casimir forces arising from thermal fluctuations. 

As already indicated in the previous section, at nonzero temperature $T$, one simply replaces the continuous integral over wave number $\kappa$ by a sum over Matsubara modes,
\begin{equation}
\int_0^\infty d\kappa \Rightarrow
\frac{2\pi k_B T}{\hbar c} {\sum_{n=0}^{\infty}}'
\hbox{\quad with \quad}
\kappa \Rightarrow \kappa_n = 
\frac{2\pi n k_B T}{\hbar c}\,.
\end{equation}
where again the prime on the sum indicates that the $n=0$ mode is counted with a weight of $1/2$.  In the limit of small $T$, the sum is well approximated by the zero-temperature integral, while for large $T$ it is dominated by the $n=0$ term.  In this form, the sum captures the combined effects of thermal and quantum fluctuations.

Because most Casimir experiments are carried out at room temperature for practical reasons, thermal effects can play an important role \cite{Brevik_2006,doi:10.1080/00107510600693683,PhysRevLett.104.040403,PhysRevLett.112.240401,bordag2009advances}.  Distance scales at which both the separation between objects can be consistently maintained and the materials act as strong reflectors of light typically correspond to optical frequencies, which then provide the dominant quantum fluctuations.  Thermal effects, however, introduce an additional length scale, which at room temperature is an order of magnitude larger.  One would expect to describe fluctuations in a metal using the Drude model for its permittivity \cite{PhysRevLett.84.4757,PhysRevE.67.056116},
\begin{equation}
\epsilon(\omega,T) = 1- \frac{\omega_p^2}{\omega(\omega + i\gamma(T))}
\end{equation}
where $\omega_p$ is the plasma frequency and $\gamma(T)$ is the relaxation frequency.
This model describes an ordinary conductor, which reflects at all nonzero frequencies but can be penetrated by a static magnetic field.  In limit where $\gamma \to 0$, the material becomes non-dissipative, described instead by the plasma model.  In this limit, we have a superconductor \cite{PhysRevA.78.062101,PhysRevB.81.195423}, which expels all magnetic fields.
The key difference in the Casimir force these models predict thus arises from the contribution of the zero-frequency transverse electric (TE) mode, which now give a discrete contribution to the Matsubara sum rather than an infinitesimal contribution to a continuous integral at zero temperature.  In the Drude model, this mode is not scattered appreciably by the metal, while in the plasma model it is perfectly reflected.

As a result, while for the plasma model one obtains the interaction free energy per unit area between parallel plates in the perfect conductor limit, for $a \ll \lambda_T$,
\begin{equation}
\frac{\cF}{\hbar c A} = -\frac{\pi^2}{720 a^3} 
-\frac{\zeta(3)}{2\pi} \left(\frac{k_B T}{\hbar c}\right)^3 
+ \frac{\pi^2 a}{45} \left(\frac{k_B T}{\hbar c}\right)^4
\hbox{\qquad [plasma model]}
\label{eqn:freeenergy}
\end{equation}
for the Drude model calculation one must subtract  
\begin{equation}
\frac{\cF_0}{A} = \frac{k_B T} {4 \pi}
\int_0^\infty k_\perp \log \left(1-e^{-2 k_\perp a}\right) dk_\perp  = 
-\frac{\zeta(3)}{16 \pi a^2} k_B T
\label{eqn:zerocontrib}
\end{equation}
to account for the missing TE zero mode contribution.

Of course, quantum fluctuations of a Drude material do not dissipate energy.  Rather, the dissipation arising from the relaxation term represents a coupling between the electromagnetic field and the material lattice at equilibrium, which is described by the imaginary-time/temperature ordered Green's function.  Given by $\epsilon^T(ic\kappa_n)$, it captures the coupled fluctuations of the electromagnetic field and the material degrees of freedom and is symmetric under time reversal, with no dissipation.  Experimental scattering data, in contrast, yield the retarded response function $\epsilon^R(\omega)$, which is asymmetric in time and does contains dissipation.  However, as shown in Ref.~\cite{PhysRevD.80.085021}, one can take advantage of the relationship $\epsilon^T(i c \kappa_n)=\epsilon^R(i c |\kappa_n|)$, as established in Ref.~\cite[p.\ 253]{Abrikosov75} and Ref.~\cite[p.\ 328]{LandauLifshitzS180}, to use the analytic continuation of information obtained about $\epsilon^R(i c |\kappa_n|)$, which is accessible to experiment, to calculate $\epsilon^T(i c \kappa_n)$, which determines the fluctuation force.

While one would expect the Drude model to provide a more accurate description of real materials with dissipation, both experimental and theoretical considerations argue in favor of the plasma model \cite{PhysRevA.65.052113,Mostepanenko:2021cbf}.   Most importantly, experiments \cite{Sushkov:2010cv,PhysRevLett.110.137401,PhysRevB.88.155413,PhysRevB.88.075402,PhysRevA.90.062115} on the whole seem to better match results obtained with the plasma model, rather than the Drude model, although the difficult task of achieving the necessary experimental precision is still in progress.  At least in the absence of impurities, a purely classical contribution to the free energy like Eq.\ (\ref{eqn:zerocontrib}) leads to conflict with the Nernst theorem for the entropy at zero temperature.  One promising possible solution to this problem involves using a  surface impedance model rather than Fresnel scattering, as shown in  Ref.~\cite{PhysRevA.67.062102}, because the key ambiguity in this mystery seems to lie in the behavior of the  surface currents associated with the zero mode.  More generally, since zero modes do not represent periodic fluctuations, their quantization frequently introduces subtleties.  In soliton physics, for example, collective quantization of zero modes restores translational and rotational symmetry, a process that also requires careful attention to dissipative terms \cite{PhysRevD.72.094015}.

\section*{Acknowledgments}

It is a pleasure to thank D.\ Gelbwaser, R.\ L.\ Jaffe,  and M.\ Kr\"uger for conversations and collaboration.  N.\ G.\ is supported in part by the National Science Foundation (NSF) through grant PHY-1820700.  M.\ K.\ is supported in part by NSF grant DMR-1708280.

\bibliographystyle{apsrev}
\bibliography{review}

\end{document}